\documentclass[preprint,12pt]{elsarticle}

\usepackage{amssymb}
\usepackage{amsmath}
\usepackage{oz}
\journal{Journal of Logic and Algebraic Methods in Programming}

\begin{document}

\begin{frontmatter}



\title{Diversity and Adjudication}


\author{Eerke A. Boiten}
\ead{E.A.Boiten@kent.ac.uk}
\address{School of Computing and Interdisciplinary Centre for Cyber Security\\
University of Kent, Canterbury, UK}

\begin{abstract}
This paper takes an axiomatic and calculational view of diversity (or ``N-version programming"), where multiple implementations
of the same specification are executed in parallel to increase dependability. The central notion is ``adjudication": once we have
multiple, potential different, outcomes, how do we come to a single result? Adjudication operators are explicitly defined and
some general properties for these explored.
\end{abstract}


\end{frontmatter}


\section{Introduction}
A long-standing technique for increasing the dependability of computer systems is redundancy. 
Rather than having just a single system (hardware or software) which runs the risk of
ending normal  operation, delivering the wrong result, or corruption by an attacker, multiple systems are used in parallel. 
If they are identical, this is called {\em replication}; if they are different implementations of the same specification
it is called {\em diversity} or {\em N-version programming} \cite{Avi77,Avi95,Laprie:1990,danielsreliable,Rom02}\footnote{References given are intended as
{\em indicative} rather than {\em representative} of the large body of research in this area over almost forty years.}.
If the multiple versions do not provide the same outcome (and there is no non-determinism involved at implementation level),
at least one of them must be exhibiting a fault. 
By looking at the different behaviours, it may be possible to conclude with certainty or with a high probability which 
system is ``wrong", and what the fault is. 
This process, of looking at outcomes of diversity and drawing some kind of conclusion from that, is called {\em adjudication} (or sometimes a ``decider" \cite{Laprie:1990}). Complex adjudication functions potentially defeat the purpose of diversity
as they introduce a risk of failing themselves.

In this paper, we will first (in Section \ref{democracy}) consider the standard model for diversity, 
where adjudication is done by majority voting, and the number of versions is the smallest value for which non-trivial majorities
exist, i.e.\ three. 
Following that, in Section \ref{resolutions} we consider more general adjudication operators and more general domains to study diversity
over. In doing so, we will encounter operators and
semantic domains which are quite familiar in Jos\'e Nuno Oliveira's working communities such as MPC and IFIP WG 2.1, ending
with a link to some of his very recent work.

Although the concepts around diversity have been well understood for some time \cite{Avi95,Laprie:1990}, this note aims to make a small
contribution by providing an abstract study of adjudication operators.

\section{Calculational Democracy}\label{democracy}
\newcommand{\MV}[3]{[ #1, #2, #3]}
To start off, we will consider {\em values}  only, as the outputs of versions -- rather than processes or even the programs delivering such values.
Thus, all we have to do is establish the properties of majority voting among three values. To do so we do need to be able to determine whether
values are equal, but no other properties on the domain of  ``values" are required.

We use an elementary notation for all of this.
\newtheorem{Notation}{Notation}
\begin{Notation} If three versions return values $a$, $b$ and $c$ respectively, we
denote the outcome of a majority vote between these three as $\MV{a}{b}{c}$.
\end{Notation}
Thus, the expression\ $\MV{2}{4}{2}$ represents a majority vote between three values, one of which is $4$ while the others
are $2$.
This represents diversity resolved by a familiar concrete adjudication operator, and we know what it means. To prepare for more abstract consideration of
such operators, let us consider the properties of this one in an abstract way. All variables will be universally quantified.

The first property is {\em unanimity}: when all versions agree on an answer,  that is then indeed {\em the} answer.
\begin{align}
\MV{a}{a}{a} & =  a
\end{align}
For majority voting, only two need to agree -- the {\em majority} property:
\begin{align}
\MV{a}{a}{b}& = a \label{maj}
\end{align}
Of course this subsumes {\em unanimity}.

The odd one out is not necessarily the last one listed, and in general the order of the values should make no difference
for the value decided on. This is
represented by the {\em permutation} properties, of which we give two that together imply the full set of possible permutations:
\begin{align}
\MV{a}{b}{c} & =  \MV{a}{c}{b}\\
\MV{a}{b}{c} & =  \MV{b}{a}{c}
\end{align}
The permutation properties, together with the failure to identify $\MV{a}{a}{b}$ and $\MV{a}{b}{b}$ suggest that
diversity works at the level of {\em bags} or multisets in the Boom hierarchy \cite{Hoogendijk:1992:PLB:648082.761188,Bunk94}.
The (commutative, associative, but not idempotent) binary operator involved would be ``$,$" if we hadn't artificially restricted ourselves for now to three versions.

Note that we do {\em not} consider the analysis of which of the three versions $a,b,c$ is at fault. This relates to the fault assumption
embedded in  N-version programming, which is that (for N=3) at most one version is faulty. Consequently, only if that assumption holds does adjudication provide
a correct result in the end. It should be clear that the ordering of the versions {\em would} make a difference if we were not
just interested in the final answer (``masking the fault"), but also in potentially pinpointing the faulty version
and fixing the fault. For example, \cite{Rom02} considers {\em recovery} of stateful versions: after one version
has been ``outvoted"  indicating a fault, can we reset its state to ensure the fault will not continue to affect future
invocations?

These properties together allow us to deduce the majority whenever there is one. But what if there isn't, i.e., what if
none of the rules above apply? We are dealing with an underspecified adjudication operator, which we can address
in several ways.
\begin{itemize}
\item We might just state that we should always pick one of the three even if there 
 is no majority -- but not fix which one. This amounts to  imposing a weakening of the majority property, which we call
the {\em choice} property:
\begin{align}
\MV{a}{b}{c} & \in   \{ a,b,c\}  \label{choice}
\end{align}
The {\em weak choice} property is characterised by this same formula, but constrained to the situations where $\MV{a}{b}{c}$ actually does
evaluate to a specific value.
\item However, for dependability the choice property may not be ideal: quietly picking one of three possible answers even when we know that at least
two of them are wrong. Instead, we may want to raise an exception, which could be modelled by including an explicit error
value $\bot$:
\begin{align}
\# \{a,b,c\} = 3 \Rightarrow
\ \MV{a}{b}{c} & =  \bot \label{error}
\end{align}
\item An entirely different alternative would be to view equality "$=$" in the majority property (\ref{maj}) as a rewriting relation from diversity terms, and
view terms that do not provide a majority as irreducible. In either case, computing with diversity gets complex when
we assume majority voting, having to deal with the partiality of the resolution operator.
\end{itemize}
We can {\em nest} diversity and adjudication -- this may occur in practice in recursive versioning \cite{Randell83,Rom02}, where subcomponents
of versioned components can be versioned themselves. The informally pitched title of this section refers to a particular
interpretation of this. A peculiar property, abused in district-based electoral systems through the practice of ``gerrymandering", is
that a nested majority vote may not return the overall majority -- and thus ``democracy" is  limited. For example, we have
\begin{equation*}
\MV{\MV{1}{1}{1}}{\MV{1}{2}{2}}{\MV{1}{2}{2}} = 2
\end{equation*}
i.e., if we adjudicate over the (in total) nine versions in multiple stages, the true majority may not be the winner.

We might also consider a world in which versioned and normal values live together. It may be possible to distribute and
maybe even lift operators from
the values world to versioned terms. For example, these seem harmless, for any binary operators $\otimes$ on values:
\begin{align}
a \otimes \MV{b}{c}{d} &= \MV{a \otimes b}{a \otimes c}{a \otimes d} \label{leftoplus}\\
\MV{b}{c}{d} \otimes a &= \MV{b \otimes a}{c \otimes a}{d \otimes a} \label{rightoplus}
\end{align}
However, for operators not injective in the versioned argument, this raises some semantic problems. Consider, for example
$\otimes$ as the maximum operator when $a$ is the largest value of all and $\{b,c,d\}$ are all different. In that case, the right
hand sides of (\ref{leftoplus}) and (\ref{rightoplus}) evaluate to a unanimity on $a$. If we adjudicate an inconclusive
majority vote as an error, as in property (\ref{error}), that reduces the adjudications in
 the left-hand sides to $\bot$, which does not contain enough information to decide whether $a$ is
the overall maximum or not, and thus the left-hand sides should evaluate to $\bot$ or some other error value. 
Assuming the {\em choice} property (\ref{choice}) instead, we would evaluate the left-hand sides to $a$ by picking one of
$\{b,c,d\}$, so this suffers no problems. In the perspective of irreducible non-majority terms, the equalities would only hold
conditionally, limited to the situation where all version outputs {\em can} be adjudicated by majority voting.

If the operator $\otimes$ is non-strict in one of its arguments, we could retain the relevant equation from (\ref{leftoplus}) and
(\ref{rightoplus}) even in the ``error"-interpretation.

We might want to generalise these equations to arbitrary functions, to allow us to write properties like
\begin{align}
f(\MV{a}{b}{c}) &= \MV{f(a)}{f(b)}{f(c)} 
\end{align}
-- but this would exhibit the same problems around non-injective functions, where
$square(\MV{-2}{2}{3}) $ is less-defined than $\MV{4}{4}{9}$.

Finally, ``lifting" of functions from values to versions is even more problematic. Intuitively, this relates to the
problem of ``fixing the choice" that arises with  non-deterministic expressions\footnote{For example, using
$\Box$ for non-deterministic choice, $(1 \Box 2) + (1 \Box 2)$
can return $3$ whereas $2*(1 \Box 2)$ cannot.}. An expression like
\[ \MV{a}{b}{c} \otimes \MV{d}{e}{f} \]
cannot in general be reduced to $\MV{a \otimes d}{b \otimes e}{c \otimes f}$, given the permutation equations. Its true value should be considering all nine different
combinations -- taking it outside our artificially limited domain! We will need to return to this once we have arbitrary
numbers of versions.
Intuitively we should be able to ``wire up" versioned computations -- but again our restricted context impacts here. 

Actually up to this point
we have not really separated versioning from adjudication, so the above expression denotes wired-up replicated {\em and adjudicated}
computations. To represent intuitive combinations, our calculus should allow for multiple or staged versions within a {\em single} adjudication operation.

\section{Different Adjudications, Different Domains}\label{resolutions}
In order to discuss a wider variety of adjudication operations, we need to drop some of the artificial restrictions in the previous section.
In particular,
\begin{itemize}
\item Adjudication and versioning should be separate operators. Bags of versions are created using an associative and commutative
binary operator which we will continue to denote as ``$,$". Majority voting, as just one specific adjudication operator,
will now be denoted $\mathbf{MV}$.
\item Using the ``$,$" operator, arbitrary numbers of versions $\geq 1$ can be created rather than restricted to $3$.
\item The domains of values we consider may have additional properties, e.g.\ they may be ordered or continuous domains.
\item We may need to consider more than just values or deterministic expressions.
\end{itemize}

For all the adjudication operators defined, we will also check whether they satisfy the relevant generalisations of the properties
(unanimity, majority, permutation, choice) defined above. 
The relevance of this is that any property that survives in all concrete instances is a candidate {\em axiom} for adjudication operators. The operators and properties are formally defined in  \ref{appform}.

\subsection{More counting}
Majority voting essentially relies on counting the occurrences of each reported outcome. In the case of three versions,
there is either an absolute majority or a three-way split. However, for larger numbers of versions, we may have the 
adjudication operator {\bf FPTP} ("first-past-the-post", as in UK parliamentary elections) where the value with the 
largest number of votes wins.

This satisfies the unanimity, majority, permutation and choice properties.

\subsection{Ordered domains}
If the domain of values has an ordering of the appropriate type, we may define the following additional
adjudication operators.
\begin{description}
\item[Upper and lower bounds] If the versioning is of academics all examining the same PhD thesis, 
the candidate may only pass if {\em all}
examiners agree on a pass. Thus, the adjudication operator in this context is the {\em greatest lower  bound} {\bf GLB}.
This satisfies the unanimity and permutation properties. Choice and weak choice may hold depending on the properties
of the ordering.
\item[Flat domain] If there is a special value $\omega$, such that $x \leq y \Leftrightarrow x=y \vee x=\omega$, then
the {\em partial least upper bound for flat domain} {\bf PLUBF} is an adjudication operator representing a sensible decision making
process. If $\omega$ represents a detectable failure of a system,
this adjudication operator represents the obvious response to this, with the following effects:
\begin{itemize}
\item it ignores failing versions, unless all are failing;
\item it does not return a value if there are multiple distinct non-failing outcomes.
\end{itemize}
In order to make it a total adjudication operator, it could be composed with another adjudication operator applied
on the bag of non-$\omega$ values; however, for this to work would require a generalisation to adjudications returning {\em bags of} values.

{\bf PLUBF} satisfies unanimity, weak choice, and permutation properties. Many majorities, however, will not win: a majority of $\omega$ can lose if
all proper outcomes are identical; if there are multiple proper values in the bag the outcome is undefined, irrespective of the
multiplicities.

\item[Median] With {\bf FPTP} we already had the mode; on a linearly ordered domain we can also have the median {\bf M}
as an adjudication operator. This satisfies unanimity, majority, permutation, and choice properties.
\end{description}

\subsection{Continuous domains}
So far, the weak choice property has held for all examples. If the domain of values is continuous, it may make more sense to pick a value
that is not one of the outputs provided by the versions \cite{283555}. For example, each version might be a sensor measuring the
same entity. Then we might return one of the following:
\begin{description}
\item[Average] The average value of all the outcomes of the versions. This satisfies the unanimity property still.
\item[\ldots with outliers removed] However, some of the outcomes may be so far outside the range of the
other measurements that they are indicative of faulty sensors, and need to be removed before averaging\footnote{The removal of outliers itself could be a generalised adjudication operator, from bags to bags of values, that can then be
used compositions with other (generalised) adjudication operators. Another one would be to remove all failures, as a generalised variant of $\mathbf{PLUBF}$.}.
\item[\ldots or with tolerances] The reason to tolerate different measured values may be in the first place
because the sensors have a (known) tolerance. One way to represent this would be to interpret
the measurements as intervals with sizes defined by the tolerances.
We may then return the intersection of the
intervals represented by all measurements with the respective tolerances as an adjudication.
\end{description}

\subsection{Beyond values}
We encountered a small problem with majority voting in the case where there is no majority: what to return? The choice
property represents the idea that whatever we do, we return one of the values under consideration. This, and the
formalisation of $\mathbf{WKCHOICE}$ in \ref{appform} as a {\em relational} adjudication operator, hint at the possibility
of {\em non-determinism}.

\begin{description}
\item[Choice] Generalising the output type of adjudication operators to potentially non-deterministic values, we could
simply return the non-\linebreak
deterministic choice of all possible values. As non-deterministic choice is idempotent, this would still satisfy the unanimity property, and in
the correct modified interpretation, also the choice property.

Considering {\em nested} adjudications would then mean also considering non-deterministic values as
{\em inputs} to adjudication. Non-deterministic choice between those would still work as an adjudication operator. 
Lifting previously discussed adjudication
operators to non-deterministic versions would be possible, returning the choice between all possible outcomes.

However, from the dependability point of view, non-deterministic choice is unsatisfactory as it means the result is unpredictable.
\item[Probability] Non-deterministic choice is idempotent, and thus removes multiplicity information from the
versions' outcomes. In other words, it effectively moves adjudication operators from operating on bags to operating
on sets of values.

This can be avoided by taking, instead of a {\em non-deterministic} choice, a {\em probabilistic} choice. This essentially
interprets the versions' outcomes as a probability distribution, where the probability of any given outcome is how often
it occurs, divided by the number of versions.

Probabilistic choice as an adjudication operator satisfies the unanimity property, as well as appropriately modified versions
of choice and majority properties.

Thus, whereas $2,2,4$ would be adjudicated to $2 \Box 4$ by non-deterministic choice, a probabilistic adjudication
of it would be\footnote{The probabilistic choice ``$a$ with a probability of $p$ and $b$ with a probability of $q$" is denoted by 
$a\strut_p\!\oplus_q b$.}  $2 \strut_{2/3}\!\oplus_{1/3} 4$, taking into account that $2$ occurred more frequently.

\item[Amplification] If we move on (again thinking of nesting) to probability distributions as inputs, we could just add
and reweight those distributions to come up with a new probability distribution. This would retain all those properties listed before.

However, there is a possibility here to do better than that. 
If we look at the inputs as distributions that have probabilities of ``wrong"
results which are strictly less than a half, we could (going back to the beginning!) use majority voting among the results
in order to {\em reduce} the chance of a ``wrong" result overall\footnote{Whereas a probability of error above a half would
increase the chance of obtaining ``error" overall, of course.}. This is called {\em amplification} in the context of
probabilistic algorithms. 

For example, if three versions all have a one in ten chance of returning an erroneous value, the chance that the majority vote
of these three is the erroneous value is reduced\footnote{Namely: all three wrong, or two wrong and one right (three options for that), so  $0.1^3 + 3*0.1^2*0.9$.} to 28 in 1000.

Thus, at this point not even the unanimity property holds: even with several identical outcomes, the final adjudicated value is
not that value! Other than permutation, we appear to have no candidate axioms left for adjudication.

More importantly, with this adjudication operator we have reached a semantic precision which allows us to express an ordering
on values (and with some extensions and liftings: on expressions, programs, processes) that represents that replication
leads to a measurable  improvement. Having an ordering, we should now be able to do refinement! But that is left for later.

This probabilistic view of adjudication relates directly to recent work by Jos{\'{e}} Nuno Oliveira \cite{JNO14}. In his work
on ``Relational Algebra for Just Good Enough Hardware", components are considered that, as well as or instead of having
design faults, can fail with a given probability. This is modelled using probabilistic relational methods, concentrating in the first
instance on sequential composition as an operator. The challenge set by our paper is whether the addition of a parallel\linebreak
(self-)composition operator with adjudication can be brought into the same rich structures of Kleisli, Mealy, Khatri, Rao, etc. 

\end{description}

\section{Related work}
The issue of replication has been covered from a formal methods perspective before. Some 20 years ago, Krishnan
developed a version of CCS with replication \cite{krishnan1994semantic}. Although it claims to be using
majority voting, in reality it uses ``first-past-the-post" with non-deterministic choice over any multiple winners.
Replication is an additional operator (different from replication in mobile calculi), with adjudication encoded directly
in the semantics using a ``seal" operator that ensures only one of the ``voted actions" can be executed.

\section*{Acknowledgement}
Thanks to Rog\'{e}rio de Lemos for linking my naive ideas about replication to some of the established work and
terminology in the dependability research area. Reviewers' constructive comments
 also helped to improve and enrich this paper.
Any remaining
misrepresentation of ideas and abuse of terminology in this paper is entirely due to me, of course.


\section*{References}
 \bibliographystyle{elsarticle-num} 
 \bibliography{replicate}


%\begin{thebibliography}{00}


%\end{thebibliography}
\appendix
\section{Formalisation of adjudication and general properties}\label{appform}
This formalisation is given in Z \cite{1200}, essentially using sets, and functions and relations (which are
also sets of pairs).
\subsection{Versions}
There is a universe of values $VALUE$. 
\[ [VALUE] \]
Adjudication operators operate on non-empty {\em bags}, which are modelled as total functions from the value domain to the 
natural numbers\footnote{The alternative is as partial functions from values to positive integers -- we model non-elements
as $0$ occurrences here.}. Applying a bag to a value is counting the number of occurrences of that value in the bag. The constraint encodes non-emptiness of the bag.
\[ BAG == \{ b: VALUE \rightarrow \mathbb{N} \zbar \ran b \neq \{ 0 \} \} \]
The elements of a bag form a set, and its size is a number, which is only defined for finite bags (definition 
to encode $size(b)= \Sigma_{x \in VALUE} b(x)$ omitted here).
\begin{axdef}
elements: BAG \rightarrow \pset VALUE\\
size: BAG \pfun \mathbb{N} 
\ST
elements(b) = \{ x: VALUE \zbar b(x)>0 \}
\end{axdef}
\subsection{Adjudications}
An adjudication operator is a relation between $BAG$ and $VALUE$. There are also functional and partial function special cases
of this:
\[ ADJOP == BAG \leftrightarrow VALUE \\
ADJOP_F == BAG \fun VALUE \\
ADJOP_{PF} == BAG \pfun VALUE
\]
Adjudication operators that do not assume an ordering on $VALUE$ are given below. They are partial: there may be no
majority, or no unique most popular choice:
\begin{axdef}
\mathbf{MV}, \mathbf{FPTP} : ADJOP_{PF}
\ST
\forall x: VALUE; b;BAG \dot \mathbf{MV}(b)=x \Leftrightarrow  2*b.x> size\ b\\
\forall x:VALUE; b:BAG \dot  \\
\mathbf{FPTP}(b)=x \Leftrightarrow \forall y:VALUE \zbar y \neq x \dot b.x > b.y 
\end{axdef}
Now assume $\leq$ is a partial ordering on $VALUE$.
\begin{axdef}
\leq : VALUE \cross VALUE
\ST
\forall x,y,z: VALUE \dot x \leq x\\
 \wedge\ (x \leq y \wedge y \leq z) \Rightarrow x \leq z\\
\wedge\ x \leq y \wedge y \leq x \Rightarrow x=y
\end{axdef}
Then we can define more adjudication operators. Unique greatest lower bounds may not exist on these minimal assumptions
on the ordering, thus $\mathbf{GLB}$ is partial.
\begin{axdef}
\mathbf{GLB} : ADJOP_{PF}
\ST
\forall b: BAG; z: value \dot\\
\mathbf{let}\ lbs=  \{ x:VALUE  \dot \forall y: VALUE \zbar b.y>0 \dot x \leq y \} 
\dot \\
\mathbf{GLB}(b)=z \Leftrightarrow z \in lbs \wedge \forall l:lbs \dot l \leq z
\end{axdef}
As the definition does not impose constraints on multiplicity of elements in $lbs$, the choice property does not generally hold
for $GLB$.

The flat domain adjudication operator can be characterised directly without using the domain's ordering:
\[ \omega: VALUE \]
\begin{axdef}
\mathbf{PLUBF}: ADJOP_{PF} 
\ST
\forall b: BAG; x:VALUE \dot \mathbf{PLUBF}(b)=x
\Leftrightarrow \\
\forall y:VALUE \zbar b.y>0 \dot y=x \vee y=\omega
\end{axdef}
The median can be defined as follows. It is partial as the ordering is not guaranteed to be linear.
\begin{axdef}
\mathbf{M}: ADJOP_{PF}
\ST
\forall b:BAG; x:VALUE \dot
\mathbf{M}(b) =x
\Leftrightarrow\\
2 * \Sigma [ y: VALUE \zbar y \leq x \dot b.y] \geq size(b) \wedge\\
2 * \Sigma [ y: VALUE \zbar x \leq y \dot b.y] \geq size(b)
\end{axdef}
This formalisation (values up from-, and values up to the median each make up at least half of all values in the bag)
makes it somewhat interesting to prove that $\mathbf{M}$ satisfies
the choice and majority properties. Unanimity is easier.
\subsection{Adjudication properties}
The potential properties of adjudication operators can be encoded as adjudication operators themselves, with
satisfaction of properties as relational inclusion -- except
for the permutation properties, which are already encoded in the bag representation.
\begin{axdef}
\mathbf{WKCHOICE}: ADJOP
\ST
\forall b:BAG; x:VALUE
\dot
b \mapsto x \in \mathbf{WKCHOICE} \Leftrightarrow b.x>0
\end{axdef}
An adjudication operator satisfies the choice property if it is included in $\mathbf{WKCHOICE}$
and is itself total.
\begin{axdef}
\mathbf{UNANIMITY}: ADJOP
\ST
\forall b:BAG; x:VALUE
\dot
b \mapsto x \in \mathbf{UNANIMITY}
\Leftrightarrow\\
\forall y:VALUE \zbar b.y=size(b) \dot y=x
\end{axdef}
Thus, an $ADJOP$ $a$ satisfies unanimity if $a \subseteq UNANIMITY$, i.e.\ it must only return results $x$
that are equal to the unanymous choice $y$ if one exists; otherwise $x$ is unconstrained.

In a similar way, the majority property is defined as an extension (Z refinement theorists might call it a
chaotic totalisation \cite{1200}) of majority voting $\mathbf{MV}$: it can return anything at all
if there is no majority.
\begin{axdef}
\mathbf{MAJ}: ADJOP
\ST
\forall b:BAG; x: VALUE
\dot
b \mapsto x \in \mathbf{MAJ} \Leftrightarrow \\
 (b \in \dom \mathbf{MV} \Rightarrow \mathbf{MV}(b)=x)
\end{axdef}

\end{document}